\documentclass[a4paper, amsfonts, amssymb, amsmath, reprint, showkeys, nofootinbib, twoside]{revtex4-1}
\usepackage[english]{babel}
\usepackage[utf8]{inputenc}
\usepackage[colorinlistoftodos, color=green!40, prependcaption]{todonotes}
\usepackage{amsthm}
\usepackage{mathtools}
\usepackage{physics}
\usepackage{xcolor}
\usepackage{graphicx}
\usepackage[left=23mm,right=13mm,top=35mm,columnsep=15pt]{geometry} 
\usepackage{adjustbox}
\usepackage{placeins}
\usepackage[T1]{fontenc}
\usepackage{lipsum}
\usepackage{csquotes}
\usepackage[pdftex, pdftitle={Article}, pdfauthor={Author}]{hyperref} % For hyperlinks in the PDF
\begin{document}
\title{Increasing the spatial bandwidth product in Light Field Microscopy with remote scanning}

\author{Aymerick Bazin}
    \affiliation{LP2N, Laboratoire Photonique Numéerique et Nanosciences, Univ. Bordeaux, F-33400 Talence, France}
    \affiliation{Institut d'Optique Graduate School $\&$ CNRS UMR 5298, F-33400 Talence, France}

\author{Amaury Badon}
    \email[Correspondence email address: ]{amaury.badon@cnrs.com}% Your name
    \affiliation{LP2N, Laboratoire Photonique Numéerique et Nanosciences, Univ. Bordeaux, F-33400 Talence, France}
    \affiliation{Institut d'Optique Graduate School $\&$ CNRS UMR 5298, F-33400 Talence, France}
    
\date{\today} % Leave empty to omit a date

\begin{abstract}
Achieving fast, large-scale volumetric imaging with micrometer resolution has been a persistent challenge in the field of biological microscopy. To address this challenge, we report an augmented version of light field microscopy, incorporating a motorized tilting mirror upstream the camera. Depending on the scanning pattern, the field of view and/or the lateral resolution can be greatly improved. Our microscope technique is simple, versatile and configured for both bright-field and epifluorescence modes. We demonstrate its performances with imaging of multi-cellular aggregates of various shape and sizes.\end{abstract}

\maketitle

\section{Introduction}

Optical microscopy has played an essential role in life sciences for studying, understanding and revealing new structures or phenomena, owing to its sub-micron spatial resolution and non-invasive nature \cite{rosenthal2009beginning}. By adding new contrast modalities, by enhancing the spatial resolution or providing the ability to target specifically areas of interest, cells can now be observed in great details routinely in numerous labs across the world \cite{murphy2012fundamentals}. These advances now make it possible to observe structures down to few nanometers or to image more and more complex and thicker samples \cite{mertz2019introduction}. In particular, the ability to image in depth while preserving biological specimens unspoiled and alive has provided an unprecedented insight into biological processes. Yet, imaging in depth requires optical sectioning capabilities to collect light only from one point or plane while rejecting light from other locations within the 3D sample of interest \cite{dunsby2003techniques}. This feature has a cost, (i) it requires instrumental complexity which leads to expensive systems, (ii) it is often time-consuming and (iii) the imaging depth is usually insufficient. For instance, optical coherence tomography or confocal microscopy both necessitate to scan a beam across the sample which makes acquisition time much longer than with bright field microscopy.

Alternatively, Light Field Microscopy (LFM) allows to simultaneously capture a large number of perspectives of the sample without any optical sectioning and to combine them to digitally reconstruct the scene in 3D \cite{curless1996volumetric,levoy2006light}. This method is fast as it requires a single acquisition, is experimentally simple but  still limited to small field of view, for instrumental reasons, and to shallow depths, due to the lack of true optical sectioning and for computational reasons \cite{simslight, wu2017light}. 
\begin{figure*}[htbp]
\centering\includegraphics[width=0.95\linewidth]{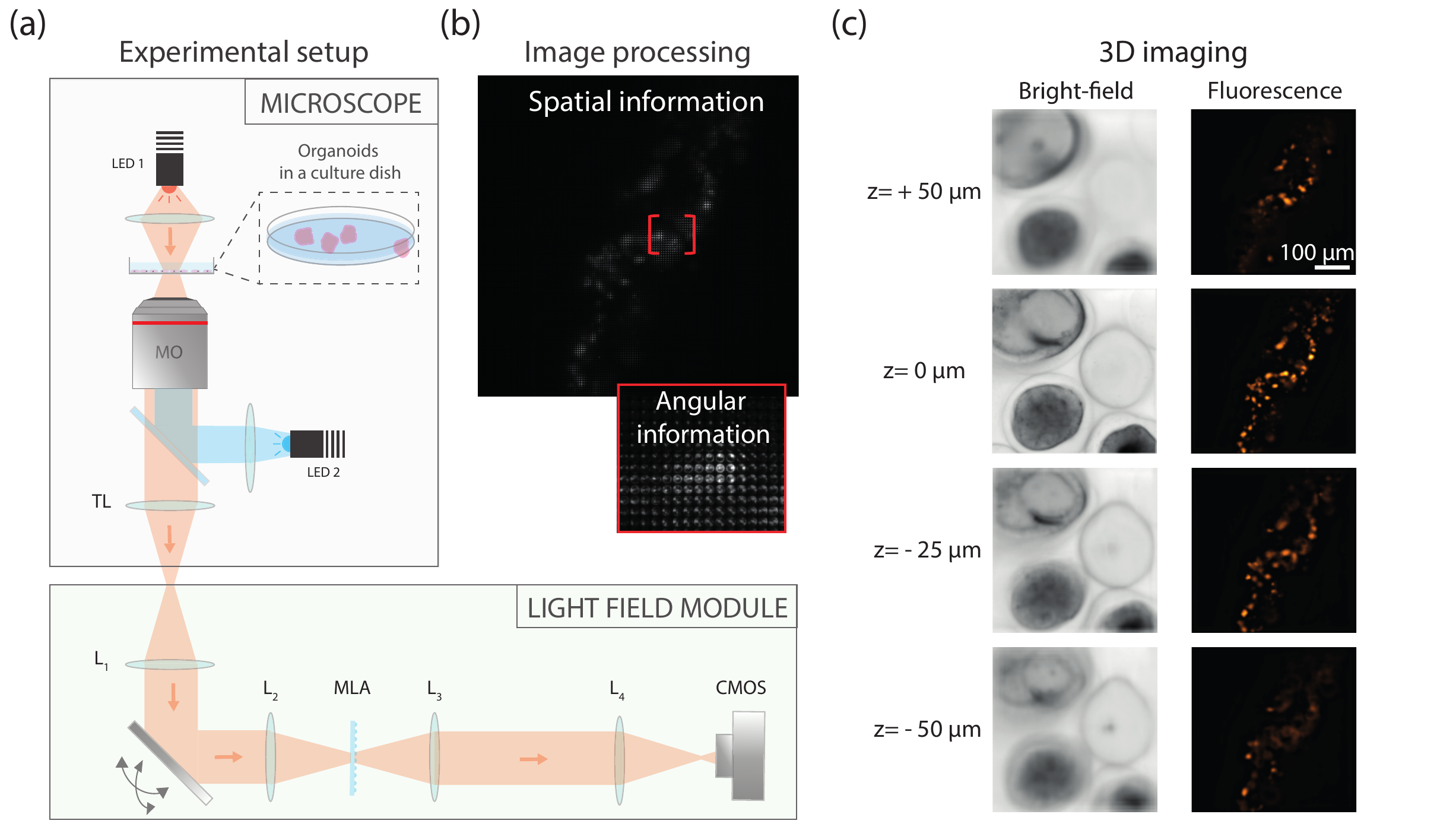}
\caption{Light field microscopy principle. (a) Experimental setup. A light field microscopy module is placed at the output port of an inverted microscope. The module is comprised of a first lens $\text{L}_1$, a motorized scanning mirror, a microlens array MLA placed in the image plane of a lens $\text{L}_2$, a relay system that conjugates the back focal plane of the MLA onto the camera sensor. (b) The camera captures raw light field images that contains both spatial and angular information. (c) Dedicated image processing allows to refocus numerically spatial information at different depths, both in bright field and epifluorescence modes. Here, we observe capsules containing human colorectal adenocarcinoma cells in bright field (left), and human dermal lymphatic endothelial cells in tubular alginate shells (right).}
\label{figure1}
\end{figure*}
Recent advances in both experimental and computational resources are driving LFM to become an increasingly powerful imaging method. For example, the growing number of pixels in CMOS sensors means that more spatial and angular information can be captured on the same sensor \cite{zhang2024long}. This means improved field of view, lateral or axial resolution. Secondly, the increased computing power of computers means that a whole volume can be reconstructed in a very short time from a single raw LF image, incorporating 3D deconvolution or deep learning methods \cite{wu2017light}. These recent improvements make it possible, for example, to image the activity of hundreds of thousands of neurons in mice, or to observe individual fluorescently-labeled organoids \cite{zhang2021computational}. Despite these recent advances, LFM remains a little-used technique due to its field of view/spatial resolution trade-off and the need to develop complex numerical methods to obtain an image as raw data are not readily interpretable.

Regarding the trade-off we face to record both angular and spatial information, the spatial bandwidth product (SBP) provides a useful insight. This quantity characterizes the throughput of an optical system. It is defined as the number of pixels necessary to capture the full field of view (FOV) at Nyquist sampling \cite{lukosz1966optical,lohmann1996space,park2021review}. As a reference point, a standard air objective 20X with a 410 nm resolution and a theoretical 1.1 $\text{mm}^2$ FOV has a SBP of 29 millions of pixels \cite{park2021review}. This quantity already exceeds by far the typical amount of pixels on a scientific camera, commonly around 4 megapixels. If now, angular information is also consider, we see that LFM with a single camera can capture only a fraction of the information transmitted by a microscope objective.

In this work, we propose a simple experimental technique for significantly increasing the SPB of an LFM, both by improving lateral spatial resolution and by drastically enlarging the lateral field of view. In this improved version, the LFM becomes a high-performance tool for imaging large numbers of multi-cellular aggregates, for example. This method relies on the use of a scanning mirror from the pupil / Fourier plane, as well as image fusing techniques. To demonstrate the performance of our approach, in this article we first present the principle of LF imaging, then show how our approach improves lateral resolution, then how to improve the lateral field of view. Next, we show how these two approaches can be combined to image a wide field of view at moderate resolution, with the possibility of increasing resolution over a region of interest. Finally, we image living multi-cell aggregates for several hours to demonstrate the performances of our approach in capturing the dynamics of living volumetric samples.
 
\section{Results}
\subsection{Principle of light field microscopy}

To obtain an improved FOV and lateral resolution, we designed an apparatus that resembles a conventional microscope body equipped with a custom made LFM detection module (see Fig.\ref{figure1}(a)). In details, a commercial microscope  (Zeiss Axiovert 200) equipped with an epifluorescence lamp (pE-2, CoolLed) is used as a base. Instead of placing directly a camera at the output port of the microscope, a tube lens $L_1$ collects the light from the native image plane and sends it toward a 2D stirring mirror (Optotune, MR-E2) placed in its the focal plane, corresponding here to the Fourier plane of the system. The microlens array, denoted MLA (Viavi, MLA-S100-f21), is then introduced in the focal plane of a second lens $L_2$ (f=200 mm) which forms an afocal system when combined with $L_1$. A second afocal system ($L_3$, f=180 mm and $L_4$, f=165 mm) re-imaged the pupil plane of the MLA onto the camera sensor (Hammamatsu, Orca Flash 4.0). 

In LFM, the 2D sensor of the camera captures both position $(x,y)$ and direction of propagation of light $(u,v)$  (see Fig. \ref{figure1}(b)). The pitch of the MLA controls lateral spatial resolution in the $(x,y)$ plane; here 100 µm pitch provides 3.1 µm in the sample plane. The focal length of individual microlenses (f=2.1 mm, NA=0.024) sets the range of angles captured. Here this range matches the NA of the objective lens divided by the magnification (see Supplementary Information for more details). Each microlens is sampled with 15 x 15 pixels to achieve satisfying angular resolution, resulting in good axial resolution. The resulting FOV at the sample is a 450 µm square with roughly 140 microlenses in each direction. 

After pre-processing steps to convert a raw light field image into a 4D light field image $I(x,y,u,v)$ (see Supplementary Information for details), independent perspectives can be extracted to observe the sample from different angles (see supplementary video 1).  Beyond this simple data manipulation, it is also possible to obtain refocused images at different depths by a combination of these perspectives. Here, we applied the so-called "shift and sum" algorithm which consists in shifting the perspectives by an amount of pixels depending on both their indices and the depth \cite{levoy2006light}. As shown in figure \ref{figure1}(c), satisfying reconstructions are obtained over a range of around 100 µm. Beyond this range, resolution and contrast are degraded but reconstruction is still doable over a 200 µm range. These reconstructions are obtained in the same way both in bright field mode and in epifluorescence mode, providing complementary information on complex biological samples. 
\begin{figure*}[ht]
\centering\includegraphics[width=0.95\linewidth]{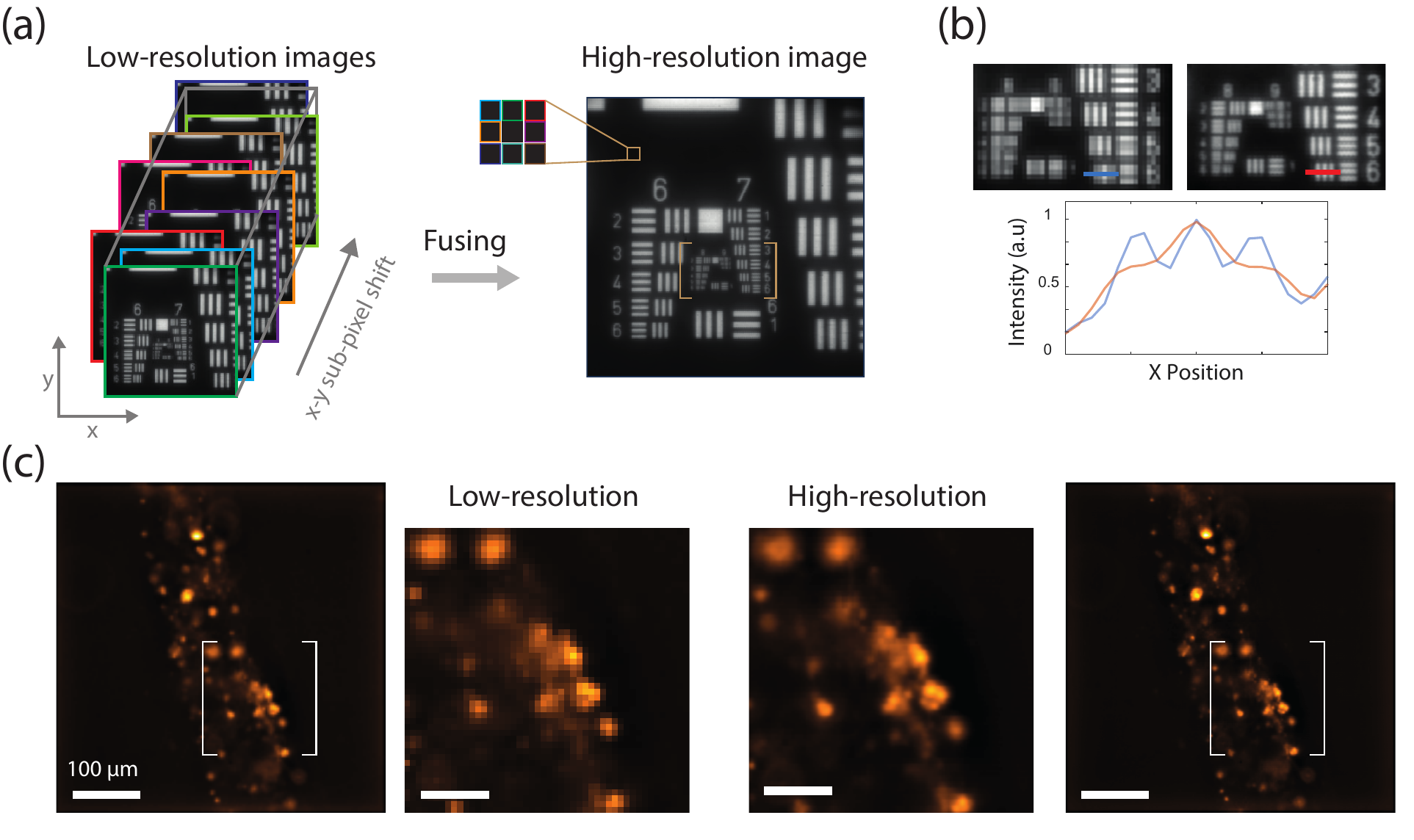}
\caption{Lateral resolution improvement. (a) Principle of the approach. A collection of 9 laterally shifted images is captured and fused into a single larger image. (b) Comparison of the intensity profiles of the same element of the USAF target (group 7 element 6) with the regular LFM and the augmented version. (c) Application on a biological sample, here human dermal lymphathic  endothelial cells encapsulated into a tubular alginate shell.}
\label{figure2}
\end{figure*}

Yet, in this configuration, both the resolution and the field of view are moderate due to the trade-off between spatial and angular sampling with a limited number of camera pixels. Our goal is to build an enhanced version of LFM with both FOV and lateral resolution improved.

\subsection{Increasing lateral resolution with remote scanning}

Lateral resolution in Light Field Microscopy is physically limited by the pitch of the microlenses, which act as macro-pixels or low-pass filters. 
Several strategies, both experimental and numerical,  have been deployed in the past to improve this lateral resolution, ideally down to the optical diffraction limit. Examples include Fourier Light Field Microscopy \cite{guo2019fourier} and numerical techniques based on aliasing \cite{broxton2013wave} or deep learning \cite{lu2023virtual}. 

Here, we adopted a multi-frame super-resolution method designed to generate a high resolution image from several low resolution images slightly shifted in X and Y \cite{farsiu2004fast,zhang2022multi}. This strategy particularly suits LFM as the mismatch between the resolution of the system and the diffraction limit is large. Implementing this technique can be done by moving either the sensor itself or the image using a tilting glass window or with a moving mirror in the pupil plane. For reasons of versatility and angular range, we installed a motorized tilting mirror in the Fourier plane of our light field detection module (see Fig.\ref{figure1}(a)). In this position, a tilt of the mirror is theoretically  solely converted into a shift in the MLA plane which is conjugated with the sample plane. Using this tilting mirror, nine images are quickly captured, each shifted by a third of a microlens in both X and Y directions (see Fig.\ref{figure2}(a)). This corresponds to a shift of 5 pixels onto the camera. After processing and reconstruction steps, the nine 3D volume (140 x 140 x 40 pixels in $x,y,z$ dimensions respectively) are fused into a single large 3D volume (420 x 420 x 40 pixels in $x,y,z$ dimensions respectively).

To demonstrate the performances of our enhanced light field microscope, we first capture a sequence of 9 slightly shifted LF images of a USAF 1951 resolution target in bright field mode. Figure \ref{figure2}(a) displays the resulting composite image which is no longer limited in resolution by the microlens pitch. A comparison of the central region clearly demonstrates a significant improvement of the lateral resolution (see Fig.\ref{figure2}(b)). Initially, only bars up to the group 5 element 6 can be resolved (4 µm) while bars up to group 6 elements 3 (1.07 µm) can now be resolved in enhanced resolution mode. This demonstrates roughly a 3 fold improvement of the lateral resolution. We note that this spatial improvement requires a sacrifice of the temporal resolution as it necessitates the acquisition of nine images plus the mirror dwell time. Typically, acquisition time is less than 3s while the fusing of the low resolution images to a larger one is almost instantaneous.

As en example demonstration, we then imaged fixed multi-cellular aggregates made of human dermal lymphatic endothelial cells in epifluorescence mode. These objects are obtained by culturing cells encapsulated within an alginate tube and serve as \textit{in vitro} models for lymphatic vessels. Thanks to a nuclear labelling (H2B-GFP), individual cells are imaged and the 3D structure can be correctly rendered. Yet, the moderate lateral resolution hinder a clear visualization especially in dense regions (see Fig. \ref{figure2}(c), left panel). If we now look at the image obtained with our method, the results are striking. Contrast and resolution are improved and we now clearly see individual nuclei with different shapes (see Supplementary Video 1).

\subsection{Remote scanning to increase the lateral FOV}

\begin{figure}[htbp]
\centering\includegraphics[width=0.95\linewidth]{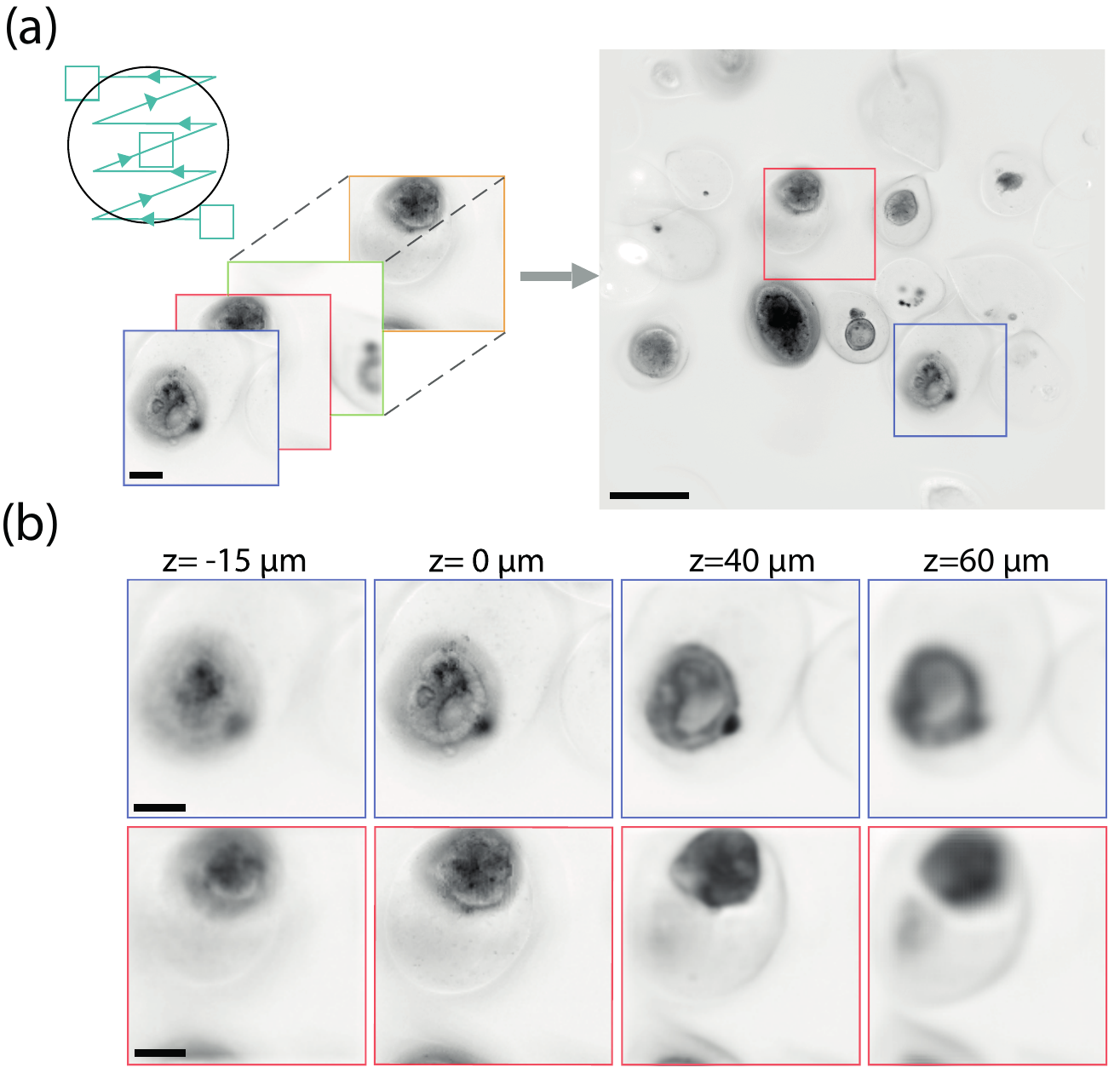}
\caption{Increased lateral field of view. (a) A collection of 25 images captured for each position of the tilting mirror is fused into a single image that covers the entire objective field of view. Here, the sample is composed of Caco2 cells encapsulated in an alginate shell. (b) Regions of interest shown in (a) reconstructed at different depths. Scale bar, 50 µm. }
\label{figure3}
\end{figure}

The enhancement of our microscope described thus far provides an increase of the SBP by simply improving the lateral resolution. Another way of increasing this quantity is to enlarge the FOV which is limited to 450 µm in our case. Here, the limitation comes from the size of the sensor that captures only a fraction of the FOV offered by the MO, which is theoretically equals to 1.1 mm for a 20X objective with an field number of 22. 

To address this issue and capture the entire FOV of the MO, denoted $FOV_{MO}$, our strategy is to sequentially redirect areas onto the sensor by using a tilting mechanism in the Fourier plane \cite{potsaid2005adaptive,recher2020remote}. Conveniently, we can use the same motorized tilting mirror as in the previous section, but this time with larger angles. To cover the entire FOV with sufficient overlap between adjacent images, required to ensure a fast and correct stitching, a sequence of 5 $\times$ 5 = 25 tilts is applied onto the motorized mirror. For each position of the sequence, a LF image is captured. In addition of  our conventional processing routine, we implemented a Fiji stitching plugin  \cite{preibisch2009globally} in Python through PyimageJ \cite{rueden2022pyimagej}. This stitching is performed onto the refocused images, ie in the spatial domain. Note that the stitching process is initially time-consuming for a calibration step, but once the coordinates are known, this procedure takes only few seconds (<2 seconds) on a conventional computer. 

Experimental validation of this approach is shown in figure \ref{figure3}(a). A population of encapsulated multi-cellular aggregates (human colorectal adenocarcinoma cells, Caco2) is imaged in bright-field mode using the 25 tilts sequence previously described. After the fusing step, the resulting 600 $\times$ 550 pixels image in x-y directions covers approximately 1.86 x 1.7 $\text{mm}^2$. Note that not all the pixels contains information as we capture a circular FOV with a squared sensor, resulting in a rectangular enlarged FOV. While a single LF barely covers a single encapsulated multi-cellular aggregate, the enlarged FOV contains a dozen of these complex biological structures. Of course, volumetric reconstruction can be performed on every area of the enlarged FOV, without noticeable reduction of the contrast or spatial resolution (see Fig.\ref{figure3}(b)). These results demonstrates that our LFM can provide large-scale volumetric imaging without moving the sample and in less than 8 seconds.

\subsection{Adaptive lateral scanning in LFM}

\begin{figure}[htbp]
\centering\includegraphics[width=0.95\linewidth]{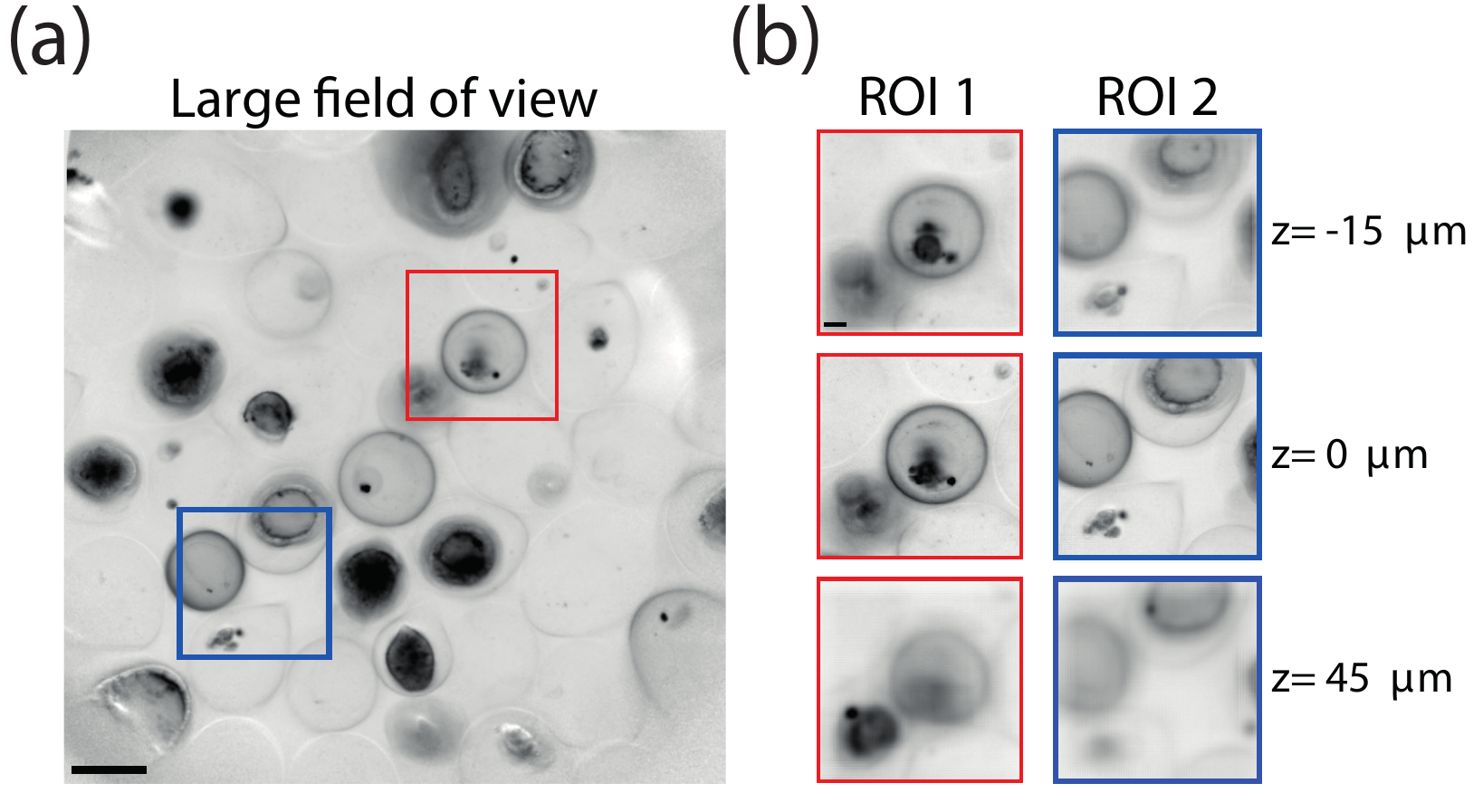}
\caption{ Adaptive scanning in LFM. (a) Large FOV obtained by stitching 25 low-resolution images of encapsulated multi-cellular aggregates. Scale bar, 200 µm. (b) Regions of interest shown in (a) imaged at higher resolution by combining multiple low-resolution images. Scale bar, 50 µm}
\label{figure4}
\end{figure}

Previously, we demonstrated the possibility to improve both the lateral resolution and the FOV at the expense of a sacrificed temporal resolution. Combining these two approaches will thus produce large-scale image at micrometer spatial resolution. In practice, our microscope can capture images that contain 1800 x 1650 pixels in $x,y$ plane, each of these spatial pixels being associated with 15 x 15 angular pixels. The time to capture this wealth of information (668 Mpixels) would be relatively short (of the order of 2 min), but more importantly it does not necessitate any motion of the sample, a critical point for fragile and alive samples in culture medium. Here, limitations would be more in terms of data quantity and processing time, in the order of dozen of minutes for a single acquisition on a conventional computer.  Based on the observation that our samples are relatively sparse, an interesting strategy will be to acquire only regions of interest (ROI) with a flexible spatial resolution. Thanks to the versatility of the motorized tilting mirror, any scanning strategy can be performed, applying large angles to cover a larger FOV or smaller angles to improve locally the lateral resolution.

As an experimental demonstration, we imaged a population of encapsulated multi-cellular aggregates, here Caco2 cells. The whole microscope objective FOV was fist captured using a 25 tilts sequence. The resulting image displays a dozens of aggregates of different sizes and shapes (see Fig.\ref{figure4}(a)). We then decided to define two ROIs to be imaged at higher resolution. Volumetric reconstruction of these two regions are displayed on figure \ref{figure4}(b) where we can clearly observe the varying shape of the aggregates as a function of depth. Such strategy is time-efficient to observe only capsules containing aggregates by avoiding empty capsules, which are in high proportion due to the microfluidic production process \cite{alessandri2013cellular}.

\begin{figure*}[htbp]
\centering\includegraphics[width=0.95\linewidth]{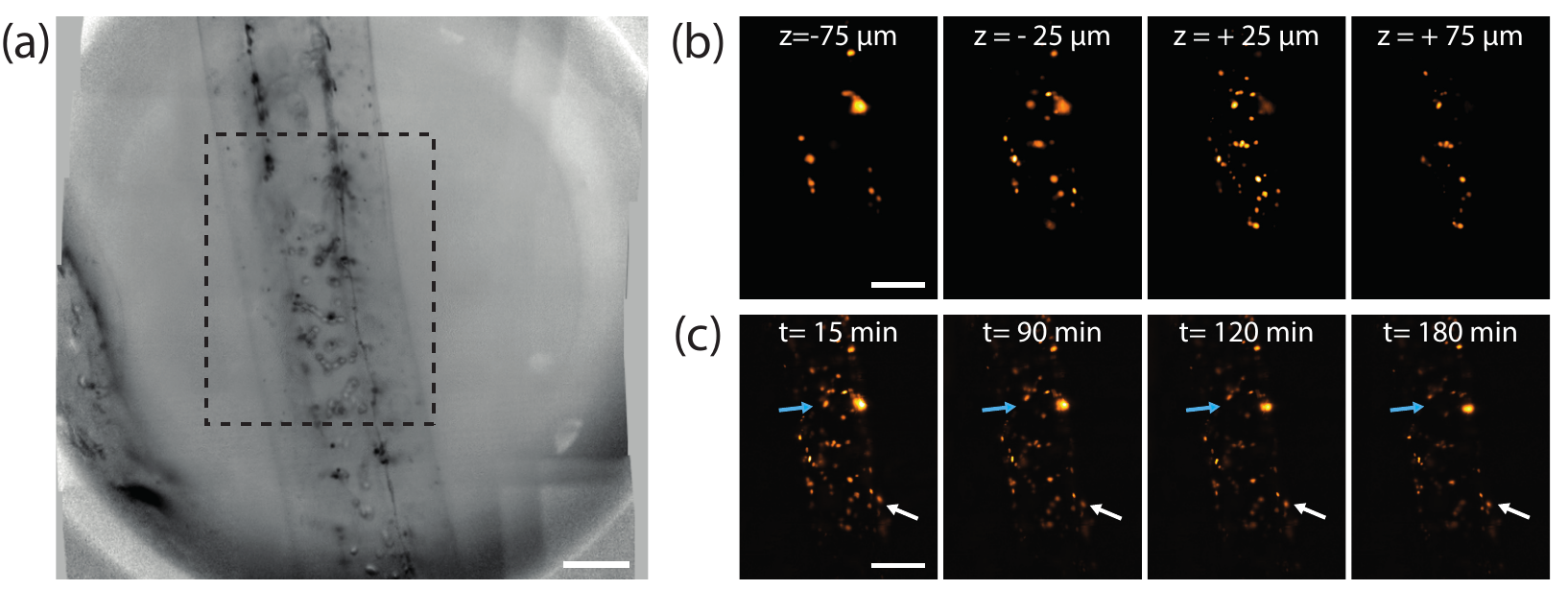}
\caption{Live imaging of encapsulated cells in an alginate tube. (a) Large FOV obtained
by stitching 25 low-resolution bright-field images. Scale bar, 200 µm. (b) Region of interest highlighted with dotted lines in (a) obtained by stitching 3 x 2 small FOV at higher resolution in epifluorescence mode at t = 0 min. Here, four different depths are displayed, showing the overall architecture of the multi-cellular aggregates. Scale bar, 100 µm. (c) Central perspective obtained of the same FOV but at different times. White and blue arrows serve as reference points to show the relative motion of cells during the temporal acquisition. Scale bar, 100 µm.}
\label{figure5}
\end{figure*}

\subsection{Live imaging}

Finally, to take advantage of the volumetric acquisition rate and low photo-toxicity of LFM, we imaged a living sample composed of human dermal lymphatic endothelial cells encapsulated in a tubular membrane. For 3 hours and every 15 minutes, we captured LFM images both in bright-field and epifluorescence modes for the whole microscope FOV and with an adapted lateral resolution. Note that acquisition rate could be greatly increased, but this is not necessary to capture the dynamics of such a sample. As seen on figure \ref{figure5}(a), the enlarged FOV offers an overall view of this multi-cellular ensemble that can span several millimeters. Then, a local observation in epifluorescence mode on a smaller region of interest provides information on the architecture of the sample. In particular, a ROI obtained by stitching 3 x 2 regular FOVs is defined from the total FOV (dotted lines in Fig. \ref{figure5}(a)). As displayed on figure \ref{figure5}(b), for t=0 minute, with a nuclear fluorescence marker, we can observe the cell arrangement inside the tube. From the median plane to the top, cells tend to be close to the alginate tube to form a hollow structure. Then, we can monitor how these cells reorganize over a period of a few hours. Blue and white arrows in figure \ref{figure5}(c) serve as guide points to highlight the displacement the cells visible here with the central perspective viewpoint. A visualization is also provided as a video (see Supplementary Video 2).

\section{Summary}

In summary, we have developed an augmented version of light field microscopy that provides non-invasive, motion-free, large scale imaging with a micrometer spatial resolution. Advantages of our system are that it is simple both experimentally and in terms of numerical approaches. First, we propose in this work to improve both the FOV and the lateral resolution with the introduction of a single motorized mirror and relay lenses, leading to an additional cost of only few k€. This is in contrast with camera array version of LFM, that offers higher performances but to the detriment of the cost of the apparatus \cite{lin2015camera}. Of course our method implies a reduction of the temporal resolution, not only because several images are needed to be captured, but also due to the settling time of the tilting mirror. This duration can be largely reduced by using faster scanning mechanism, such as galvanometer systems. Yet such systems are usually costly, achieving large angle tilts at high speed is challenging especially with extended beam diameters, which require large mirror size. \\

Second advantage is that no complex numerical approach is involved in our work. Our method to increase the FOV only necessitates to stitch images according to the pattern applied on the scanning mechanism. Here, we adapted a well-known, high performance stitching algorithm \cite{preibisch2009globally}. As it stands, it doesn't seem possible to make any changes to speed up its operation, it would require developing a a whole new architecture from scratch. 
Regarding the lateral resolution improvement technique, we have opted for a simple solution involving the manipulation of pixel indices between images. Computation time is therefore extremely short. More powerful approaches that consider blurring and warping, such as deconvolution \cite{ben2005video} or deep learning \cite{zhang2022multi,xue2022deep}, can probably further improve lateral resolution at the cost of longer computation times. \\

Finally, despite its experimental simplicity, the footprint of our LFM detection module is significant, meaning it requires part of an optical table to be mounted on. In order to facilitate the use of such a system, we plan to develop a compact version of this module, by suppressing the second relay system and add the MLA directly in front of the sensor. To go even further, a compact version of the whole system could be developped in order to fit inside a table top incubator, which would offer an \textit{in situ} observation of live specimen over days or weeks in ideal conditions \cite{badon2022incubascope,jana2024zincubascope}. 

The simplicity and ease of use of our system should make it attractive for general biomedical research applications.

\subsection*{Funding}
The authors acknowledge the financial support from the French National Agency for Research (ANR-22-CE42-0019) and GPR Light.

\subsection*{Acknowledgments}

We would like to thank all the BiOf team for fruitful discussions and especially Camille Douillet, Elsa Mazari-Arrighi and Adeline Boyreau for providing biological samples.

\subsection*{Disclosures}

The authors declare no conflicts of interest.

\subsection*{Data Availability Statement}

Data underlying the results presented in this paper are not publicly available at this time but may be obtained from the authors upon reasonable request.

\subsection*{Supplemental document}
See Supplement 1 for supporting contents.

%%%%%%%%%%%%%%%%%%%%%%% References %%%%%%%%%%%%%%%%%%%%%%%%%

%%%%%%%%%% If using BibTeX:
\bibliography{LFM_arxiv}

\appendix*

\end{document}